\documentclass[aps,prl,twocolumn,showpacs,floatfix,10pt]{revtex4}
\usepackage{amsmath,graphicx,subfigure}

\newcommand{\qcOM}{q_{\textrm c,{\textrm{OM}}}^2}
\newcommand{\IthOM}{I_{\textrm{th}}^{{\textrm{OM}}}}
\newcommand{\qc}{q_{\textrm{c}}^2}
\newcommand{\Ith}{I_{\textrm{th}}}
\begin{document}
\title{Opto-Mechanical Pattern Formation in Cold Atoms}
\author{E. Tesio}
\author{G.R.M. Robb}
\author{T. Ackemann}
\author{W.J. Firth}
\author{G.-L. Oppo}
\affiliation{ICS, SUPA and Department of Physics, University of Strathclyde, Glasgow, G4 0NG, U.K.}
\date{\today}
\begin{abstract}
Transverse pattern formation in an optical cavity containing a cloud of cold two-level atoms is discussed. We show that density modulation becomes the
dominant mechanism as the atomic temperature is reduced. Indeed, for low but achievable temperatures the internal degrees of freedom of the atoms can be neglected, and the system is well described by treating them as mobile dielectric particles.
A linear stability analysis predicts the instability threshold and
the spatial scale of the emergent pattern. Numerical simulations in one and two
transverse dimensions confirm the instability and
predict honeycomb and hexagonal density structures, respectively,  for the blue and red detuned cases.
\end{abstract}
\maketitle
% ======= INTRODUCTION =======
Pattern formation can be defined as the spontaneous emergence of spatio-temporal
structures in nonlinear systems driven far from equilibrium~\cite{CrossHoh}. The spatial
structure emerges from an initially homogeneous state as a consequence of the interplay
between local nonlinearity and spatial mechanisms such as diffusion and diffraction. In
optical systems, spatial structures of the light intensity are generated in the plane
transverse to the direction of propagation after interaction with a nonlinear medium in
the presence of feedback. Several geometries have been shown to provide
the required feedback: two counterpropagating beams~\cite{Gry:OptComm:88},
ring or Fabry-P\'erot cavities~\cite{Lug:PRL:87,Lug:PRA:88,Lippi:CSF:94}, or the single
mirror feedback arrangement~\cite{giusfredi88,Firth:JMO:90,Dalessandro:PRL:91,Ackemann:PRA:94}. Moreover, theoretical
and experimental investigation demonstrated that different media can support the formation
of transverse structures, such as (hot) atomic gases~\cite{Gry:OptComm:88,giusfredi88,Lippi:CSF:94, Ackemann:PRA:94},
Liquid Crystal Light Valves~\cite{Neubecker:PRA:95} or photorefractive media~\cite{Denz:Book:03}.
In particular, atomic media present the appealing feature of being describable from first principles
combining the microscopic equations for the medium and the Maxwell equations for the
incident and generated radiation. It is well known that diffraction and optical nonlinearities can induce
spatio-temporal structures inside the medium through modulation of populations and coherences.\\
An additional mechanism for spontaneous self-organization appears in atomic
media cold enough for optical forces to compete with thermal effects sufficiently strongly to modify the density distribution of the sample.
Nonlinear optical effects involving the mechanical effect of light on cold atoms are well known and
have been predicted and demonstrated in cold atomic gases~\cite{Bonifacio:NIMA:94,vonCube:PRL:04,
Slama:PRL:07} and Bose-Einstein Condensates~\cite{Inouye:Science:99,Schneble:Science:03}.
In particular, the possibility of using such effects for spontaneous filamention has been proposed in~\cite{Saffman:PRL:98}.
The nonlinear interplay between these opto-mechanical forces and the induced density modulations that they create
and respond to has been shown to give rise to collective or cooperative light scattering~\cite{Bonifacio:NIMA:94,vonCube:PRL:04,Slama:PRL:07,Inouye:Science:99,Schneble:Science:03,Robb:PRL:03,Wiggins:JMO:02,adam03,baumann10} in atoms and self-focusing~\cite{Ashkin:OptLett:82},
and four-wave mixing~\cite{Smith:OptLett:81} in solutions of dielectric spheres.
However, the formation of opto-mechanical structures, arising from spontaneous symmetry breaking in the plane orthogonal to the pump axis due to opto-mechanical forces, has not been addressed in these studies, and is the main topic of this Letter.
For counterpropagating beams, a lowering in the threshold for
transverse self-organisation on the focusing side of the nonlinearity has been proposed~\cite{Muradyan:ThB29:2005,Saffman:LNP:2008}. Experimental evidence of the formation of transverse
structures in cold atomic media has also been found in~\cite{Greenberg:OptExpr:11,Guerin:PRL:08}.
These previous studies typically  emphasize the interplay between mechanical density redistribution
effects and the nonlinearities arising from the internal degrees of freedom of the atom, and often involve multiple optical beams forming wavelength-scale lattices, and perhaps also optical polarization effects related to the multi-level quantum structure of the atoms.\\
In contrast, we consider a very simple, and hence general, system, of ground-state atoms interacting
with a single coherent optical field. The linear dielectric response of the atoms, which is responsible
for the refractive index of the cloud, means that the atoms will move up or down any transverse gradient
in the optical field. In turn, the refractive effects of non-uniform atomic density will lead to phase gradients, and thus eventually intensity non-uniformities, in the optical field. As we will show, this simple mechanism
readily produces positive feedback, and thus transverse instability, independently of the the sign of the atomic response. We illustrate this very general instability mechanism for the case of a Doppler cooled two-level atom cloud within a planar ring cavity driven by a monochromatic plane wave optical field.
This simple fundamental atom-field coupling is, of course, already present in all the above-mentioned experiments, and in many of the models. Its importance has not previously been apparent because of
the complexity of the systems considered.  We believe that the results of our simple model will be important for the
interpretation of these more complex experiments and models, and hence for future progress in cold-atom optics.
% *************************
%In many of these studies, the most significant material quantity is not the
%spatial distribution of coherences or populations but instead the spatial distribution of atomic or
%particle density which is typically driven by optical gradients or dipole forces.
%and suspensions
%of dielectric particles \cite{Ashkin:OptLett:82,Smith:OptLett:81,Robb:PRL:03,Wiggins:JMO:02}.
% ************ SETUP FIGURE
\begin{figure}
 \centering%
 \includegraphics[scale=0.4]{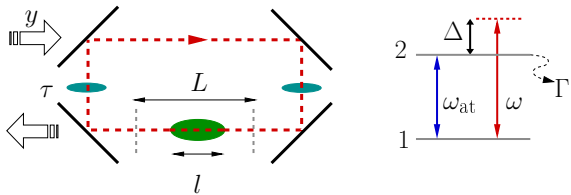}
 \caption{ (color online) A sample of two level atoms with thickness $l$ and density $N_0$ is inserted in a
 planar ring cavity of effective length $L$ and laser-cooled at a temperature $T$. $L$ can be controlled
 by adjustment of the intra-cavity lenses, and is therefore distinct from the pysical cavity length.
 A plane wave of amplitude $y$ and frequency $\omega$, detuned from the resonance $\omega_{at}$
 by $\Delta$, pumps the medium in the cavity. The transmittivity of the mirrors is $\tau$. A
 typical configuration can be obtained exploiting the $D_2$ line ($\lambda=780$ nm) in a sample
 of $^{85}$Rb of thickness $l=10\,$mm and density $N_0=7\times 10^{10}$at/cm$^3$, cooled
 to twice the Doppler temperature ($T_D=150\,\mu$K) and inserted in a cavity with mirror
 transmittivity $\tau=0.1$.}
 \label{fig:cavity}
\end{figure}
% *************************

We consider a sample of $N$ identical, non interacting two-level atoms inserted in a planar ring
cavity and cooled to temperature $T$ (see Fig.~\ref{fig:cavity}). We allow for a transverse
redistribution of the sample density under opto-mechanical forces in the form
$N(\mathbf x,t)=N_0 n(\mathbf x,t)$, where $N_0$ is the average density of the sample and
$n(\mathbf x,t)$ encodes a spatial density modulation. Such a modulation will enter the expression
for the sample susceptibility, with regions of higher density corresponding to larger responses
of the medium to the incoming radiation. The material susceptibility is cast in the form
$\chi=N(\mathbf x,t)\chi_{\textrm e}(\mathbf x,t)$, where $N$ accounts for the density redistribution
effects and $\chi_\textrm e$ represents the electronic susceptibility of the medium. Adiabatic
elimination of populations and coherences in the optical Bloch equations of a two-level system
are known to produce an intensity-dependent electronic response of the form
$\chi_\textrm e=\chi_\textrm e(|f(\mathbf x,t)|^2)$ where $f$ represents the amplitude of the
electric field~\cite{Lug:PRA:88}. The medium susceptibility $\chi$ then acts as a nonlinear source
term in the field wave equation, which can be written in the slowly varying envelope, rotating
wave, paraxial and mean field approximations as~\cite{Lug:PRA:88}:
\begin{equation}
 \dot f = -(1+i\theta)f +y -\tilde\gamma\,n\frac{f}{1+\frac{|f|^2}{1+\Delta^2}} + i\nabla^2_\perp f\,.
\label{eq:field}
\end{equation}
Eq.~(\ref{eq:field}) contains only adimensional quantities, with the field $f$ rescaled to the
saturation intensity at resonance and the time normalized to the cavity losses (see the term
$-f$). $\Delta$ is the light-atom detuning (in units of the coherence decay rate $\Gamma/2$), while
$\nabla^2_\perp=\partial_x^2+\partial_y^2$ denotes the transverse Laplacian and describes diffraction.
The cavity imposes a linear phase shift governed by the cavity detuning $\theta$, and is pumped by
a plane wave of (normalized) amplitude $y$. Definitions are chosen so that $\Delta>0$ ($\Delta<0$)
corresponds to blue (red) detuned beams and thus to self-focusing (defocusing) nonlinearities.
The strength of the complex susceptibility is $\tilde\gamma=2C(1+i\Delta)/(1+\Delta^2)$, with absorption and dispersion captured
by its real and imaginary parts, respectively. The cooperativity parameter $C$ contains the dependence
on the sample density $N_0$ and the mirror transmittivity $\tau$. Spatial coordinates are normalized
to the diffraction length $\sqrt a=\sqrt{\lambda L/4\pi \tau}$, where $\lambda$ is the radiation
wavelength and $L$ the effective cavity length (see Fig.~\ref{fig:cavity}). Note that terms in the
Maxwell equations varying as $\dot n$, $\ddot n$ are neglected in deriving Eq.~(\ref{eq:field}).\\
If $n$ is considered uniform, Eq.~(\ref{eq:field}) becomes a standard model of cavity nonlinear optics,
showing optical bistability and pattern formation for appropriate parameter choices.
If the nonlinear term in Eq.~(\ref{eq:field}) is neglected, no instability is possible for constant $n$, and the role of the atoms is simply to modify the cavity losses and resonance frequencies through the complex
linear susceptibility. If we allow the density to enter as a dynamical variable, however, the system behavior becomes qualitatively different. To proceed,  it is necessary to specify an equation for the dynamics of the density modulation $n$, which is coupled
back to the field through the action of optical forces. In the limit of large detuning scattering
forces are negligible, and the sample is subject only to a conservative dipole potential
$U_{\textrm{dip}}=(\hbar \Gamma\Delta|f|^2) / 4(1+\Delta^2+|f|^2)$. Assuming a strong viscous
damping of the momentum distribution, e.g., due to the presence of an optical molasses, a Fokker-Planck equation can be derived for the dynamics of
the atomic density~\cite{Saffman:LNP:2008,OHara:PRA:01}:
\begin{align}
 \dot n= \sigma D\nabla_\perp\cdot&\left[n\nabla_\perp \frac{|f|^2}{1+\Delta^2+|f|^2}\right]+
 D\nabla_\perp^2 n\label{eq:density}\,,\\
 &\hspace*{1cm}\sigma=\frac{\hbar\Gamma\Delta}{4k_\textrm{B}T}\,.\nonumber
\end{align}
We recognise the first term on the
right side of Eq.~(\ref{eq:density}) as the divergence of a drift current originating from the transverse dipole forces, potentially leading to non-uniform density, while the second term, diffusion, opposes  such non-uniformities. The
parameter $\sigma$ characterizes the relative strength of these opposing tendencies. Clearly density
modulation is favored by large detuning ($\Delta$) and, importantly, by low temperature. In the limit of high temperatures ($\sigma\to 0$) diffusion drives the atomic
distribution towards the homogeneous state ($n=1$), and the model reproduces the results corresponding to hot atomic vapours.
The stationary state for the density  modulation is given by the equilibrium distribution
\cite{Saffman:LNP:2008,OHara:PRA:01}:
\begin{equation}
 n_{\textrm{eq}}(\mathbf x)=\frac{V\exp\left(-U_{\textrm{dip}}/k_\textrm{B}T\right)}
 {\int_Vd\mathbf x\exp\left(-U_{\textrm{dip}}/k_\textrm{B}T\right)}\,.
\label{eq:Gibbs}
\end{equation}
% ================================
\par
\begin{figure}
 \centering%
 \includegraphics[scale=0.4]{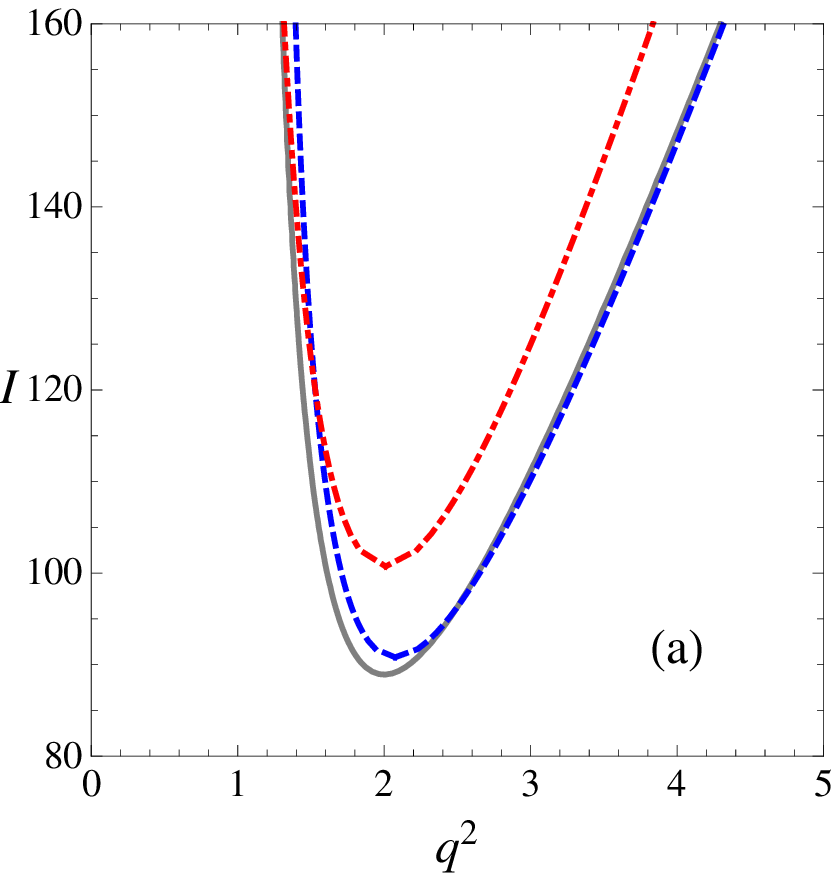}\quad
 \includegraphics[scale=0.4]{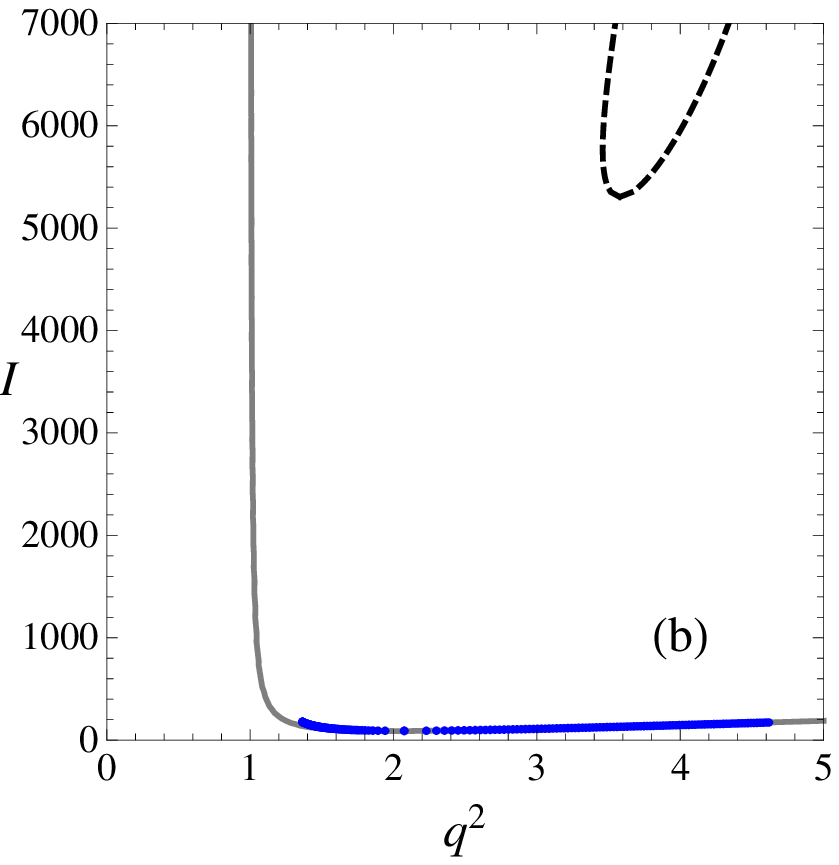}
 \caption{(color online) Left: %in text, refer to "a" and "b", not left and right
 pattern-formation threshold curves for Eqs.~(\ref{eq:pureOM}) (full line) and for the full system
 Eqs.~(\ref{eq:field},\ref{eq:density}) for blue (dashed line) and red (dash-dotted line) detuning.
 $|\Delta|=100$, $|\sigma|=25$,
 $C=225$ and $\theta+\gamma=-1$. Right: on a much larger intensity scale,
 thresholds for the purely mechanical system (full line), the complete system (blue dots) and the
 blue-detuned case of the saturable Kerr model (dashed line) without density modulation.}
 \label{fig:threshold}
\end{figure}

% ======== PURE OM MODEL =========
Eqs.~(\ref{eq:field}) and~(\ref{eq:density}) describe the coupled dynamics of the intra-cavity field
and the two-level sample when both electronic and opto-mechanical effects are present. This kind of
system, and the role played by the temperature, has been addressed for the arrangement of two
counterpropagating beams in~\cite{Muradyan:ThB29:2005,Saffman:LNP:2008}. Here we study the much simpler situation of a unidirectional beam and, moreover, negligible electronic nonlinearity. We show that opto-mechanical effects alone are capable of providing a pattern-forming instability,  through density redistribution. For large detuning ($|\Delta|\gg1$) scattering forces and absorption are negligible compared to dipole forces and dispersion, respectively. If we also neglect the  electronic nonlinearity, the system
(\ref{eq:field}-\ref{eq:density}) reduces to
\begin{subequations}
 \begin{align}
  \dot f&=-(1+i\theta)f+y-i\gamma\,n\,f+i\nabla^2_\perp f\label{eq:pureOM:field}\\
  \dot n&=\sigma D\nabla_\perp\cdot\left[n\nabla_\perp\frac{|f|^2}{1+\Delta^2}\right]+ D\nabla_\perp^2 n
  \label{eq:pureOM;density}
 \end{align}
\label{eq:pureOM}
\end{subequations}
where $\gamma=\hbox{Im}[\tilde\gamma]=2C\Delta/(1+\Delta^2)$ accounts for linear dispersion and
nonlinear terms in $|f|^2/(1+\Delta^2)$ have been neglected. We remark that this limit can be
experimentally  feasible. For a $10$ mm thick sample of $^{85}$Rb with density
$N_0=7\times 10^{10}$ at/cm$^3$ at a temperature $T=300\,\mu$K, interacting with a laser beam
detuned by $|\Delta|=100$ linewidths from the $D_2$ line and mirror transmittivity of $10\%$,
for instance, one obtains $C\simeq 225$ and $|\sigma|\simeq25$, and we will find that the electronic
nonlinearity is indeed small at the threshold for density-driven pattern formation.
% ====== LIN. STAB. ANALYSIS ======
To demonstrate this, we perform a linear stability analysis of the system (\ref{eq:pureOM}). The flat,
stationary state of (\ref{eq:pureOM}) is given by $f_\textrm s=y\left[1+i(\theta+\gamma)\right]^{-1}$,
$n_s=1$. We perturb this flat solution as $f=f_\textrm s+\delta f(\mathbf x,t)$,
$n=1+\delta n(\mathbf x,t)$ and linearize the system (\ref{eq:pureOM}). When looking for static
instabilities we write the perturbations in the form
$\sim e^{i\mathbf q\cdot \mathbf x_\perp}e^{\lambda(\mathbf q) t}$ (with $\lambda$ real) and impose
the condition of marginal stability, i.e.\ $\lambda(\mathbf q)=0$. A threshold condition is found for
the control parameter $I=|f_\textrm s|^2$:
\begin{subequations}
\begin{align}
 &\qcOM=1-(\theta+\gamma)\label{eq:threshold:qc}\,,\\
 &\IthOM=%\frac{(1+\Delta^2)\left[1+(\theta+\gamma+\qcOM)^2\right]}{2\sigma\gamma(\theta+\gamma+\qcOM)}=
         \frac{1+\Delta^2}{\sigma\gamma}\,.
\label{eq:threshold:I}
\end{align}
\label{eq:threshold}
\end{subequations}
Here $\mathbf q_{\textrm c,{\textrm{OM}}}$ denotes the critical wavenumber of the system
(\ref{eq:pureOM}), i.e.\ the first transverse mode to become excited when increasing the control
parameter. The critical wavenumber can be controlled by varying the cavity detuning $\theta$:
we set $\theta+\gamma=-1$ ($\theta\simeq-5.5$), which gives $\qcOM=2$. $\IthOM$ represents the minimum value of intensity such that the growth rate crosses
zero, $\lambda(\mathbf q_{\textrm c,{\textrm{OM}}})=0$. We find  $\IthOM\simeq88.9$ for our
choice of parameters. The low-excitation assumption is therefore confirmed since
$\IthOM/(1+\Delta^2)\simeq0.009\ll 1$.\\
Figure~\ref{fig:threshold} shows the threshold curves $I(q^2)$ for our choice of parameters.
% **************************
% ********* THRESHOLD FIGURE
% **************************
Fig.~\ref{fig:threshold}a compares the threshold for the purely opto-mechanical system of
Eqs.~(\ref{eq:pureOM}) (full line) with that of the full system involving both electronic and mechanical
effects, Eqs.~(\ref{eq:field}-\ref{eq:density}), for blue (blue dashed line) and red (red dot-dashed
line) detuning. Electronic nonlinear effects clearly  cause only small corrections to the threshold (\ref{eq:threshold}) from the purely opto-mechanical model. The minimum thresholds for the full
model are found
to be $\Ith\simeq 90.8$ (blue detuning) and $\Ith\simeq 100.7$ (red detuning). The critical
wavenumbers are $\qc\simeq 2.07$ (blue detuning) and $\qc\simeq 2.01$ (red detuning).
The instability behavior of the full system is clearly dominated by the opto-mechanical modulation
of the atomic density, with electronic effects negligible in first approximation. Further evidence is
displayed in Fig.~\ref{fig:threshold}b, which compares the thresholds for the purely mechanical system of
Eqs.~(\ref{eq:pureOM}) (full line) with that for a saturable Kerr medium
with no density redistribution effects, i.e.\ a hot two-level medium (dashed line), which is more
than two orders of magnitude greater (for blue detuning, $\Delta=100$: there is no hot-atom
instability for red detuning for our choice of $\theta$).

We note that instability thresholds arising
from opto-mechanical effects are independent of the sign of the detuning, as a change in the sign
of $\Delta$ (and thus $\gamma$) is compensated by a change in the sign of $\sigma$. This is strongly
reminiscent of the artificial Kerr media studied by Ashkin and collaborators~\cite{Ashkin:OptLett:82},
where dielectric particles are subject to light forces. A change in the sign of the particle
polarisability (i.e.\ from blue-detuned to red-detuned beams) is compensated by the fact that
positively (negatively) polarised particles are pushed to the maxima (minima) of the field intensity.
The value of the temperature is crucial in determining the interplay between the opto-mechanical and
the electronic nonlinear mechanisms. The parameter $\sigma$ represents in fact the ratio between the
dipole energy $\hbar(\omega-\omega_{\textrm{at}})$ and the thermal energy $k_BT$. By increasing this
ratio, opto-mechanical nonlinear effects arising from dipole forces become dominant over the electronic
nonlinearities from saturation of the involved transition.
Internal degrees of freedom thus become
negligible close to the instability threshold which can thus be interpreted as due to opto-mechanical
effects alone (see Fig.~\ref{fig:threshold}). We remark that, although $|\sigma|$ can, in principle,
always be increased to very high values by lowering the temperature, our parameters
do not require sub-Doppler temperatures, because our large detuning parameter weakens the
electronic reponse relative to the opto-mechanical.

\begin{figure}
 %\centering%
 \includegraphics[scale=0.21]{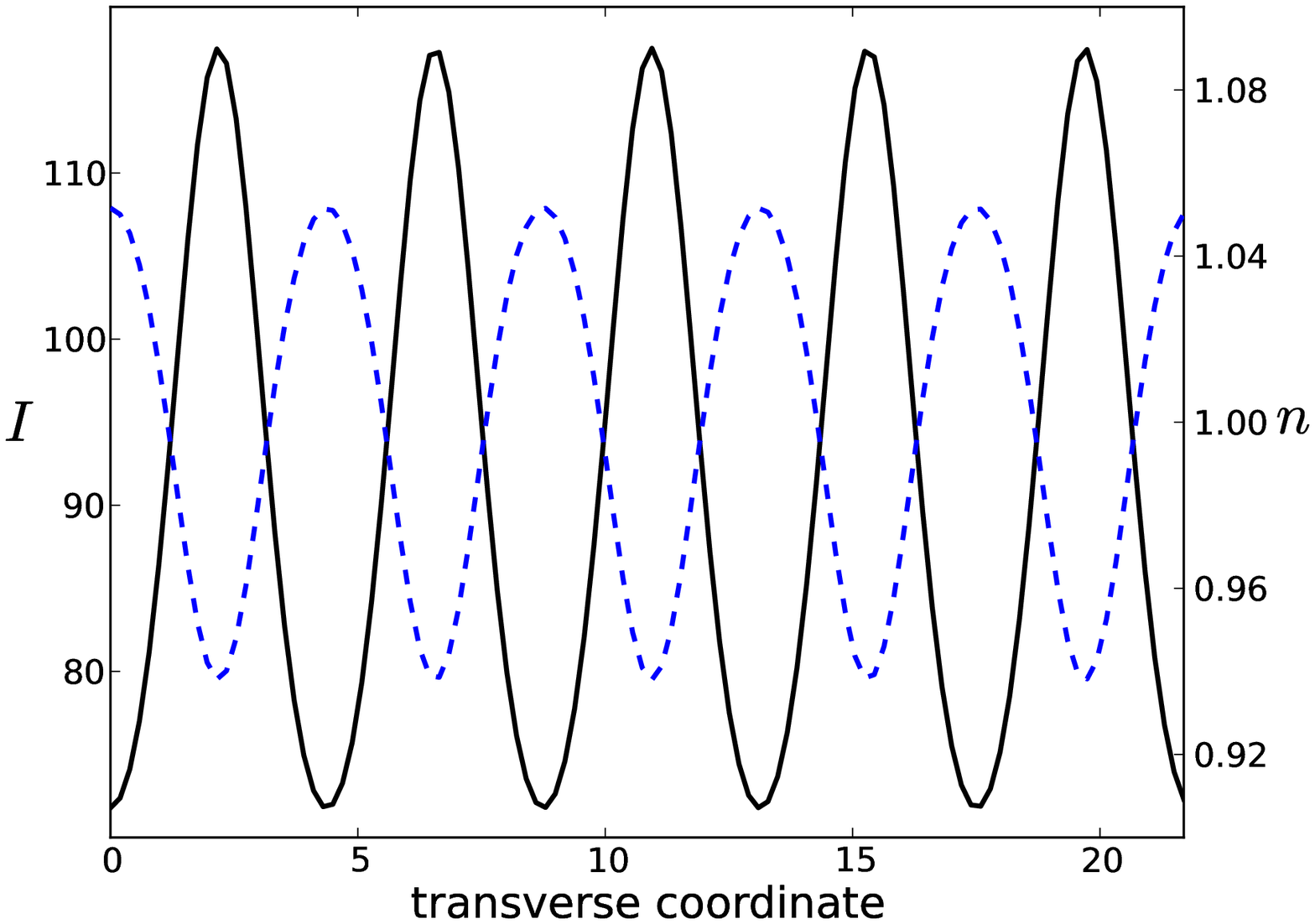}
 \includegraphics[scale=0.21]{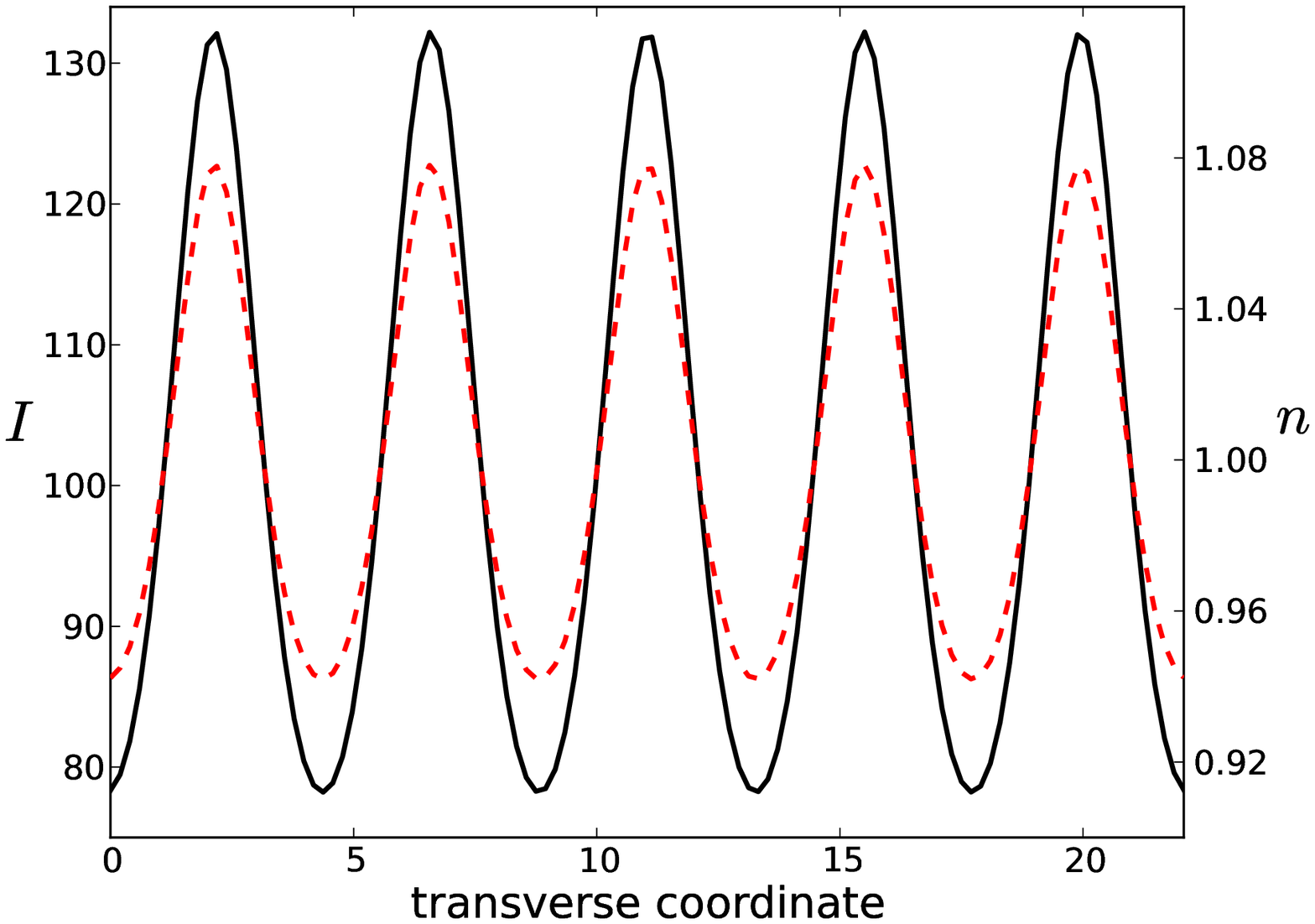}
 \caption{ (color online) Intensity (black solid lines) and density (dashed lines) transverse structures in one
 transverse dimension obtained from numerical simulations of the system~(\ref{eq:field}-\ref{eq:density})
 at about $3\%$ above threshold. Parameters are as in Fig.~\ref{fig:threshold}, with the blue-detuned
 case on the left panel and the red detuned case on the right panel. Electronic nonlinear effects can be
 neglected, so that the system undergoes a pattern-forming bifurcation with critical wavenumber
 $\qc\simeq2.07$ ($\qc=2.01$) and threshold intensity $\Ith=90.8$ ($\Ith=100.7$) in the blue (red)
 detuned case. Density structures form according to the values of detuning and
 temperature, see Eq.~(\ref{eq:Gibbs}).}
 \label{fig:simulations}
\end{figure}
\par
% ==================================
% ======== NUM. SIMULATIONS ========
Numerical simulations in one and two transverse dimensions have been performed to test
the predictions obtained from the linear analysis.\ The models (\ref{eq:field}-\ref{eq:density}) and (\ref{eq:pureOM}) have been integrated using a second-order Crank-Nicholson method, with the density dynamics given by
(\ref{eq:Gibbs}).\ Periodic boundary conditions are imposed over a domain of $5$ critical wavelengths
$\lambda_c=2\pi/q_c$.\ The transverse domain is discretized using $112$ points for one-dimensional (1D) simulations, and a
square grid of $80\times80$ points for two-dimensional (2D) simulations.\ Time is discretized with
step $dt=5\times10^{-3}$.
Fig.~\ref{fig:simulations} presents results of 1D numerical integration of the
full system ~(\ref{eq:field}-\ref{eq:density})
for the parameters of Fig.~\ref{fig:threshold} and intensities slightly above threshold, $I=92.6$
(blue detuning) and $I=103.7$ (red detuning). In Fig.~\ref{fig:simulations}
the dashed lines correspond to the density distribution: from Eq.~(\ref{eq:Gibbs}) the absolute
value of $\sigma$ governs the amplitude of the modulation, while its sign determines
whether atoms bunch in regions of high ($\sigma<0$) or low ($\sigma>0$) intensity, see
Fig.~\ref{fig:simulations}.
\begin{figure}
 \centering%
 \includegraphics[scale=0.17]{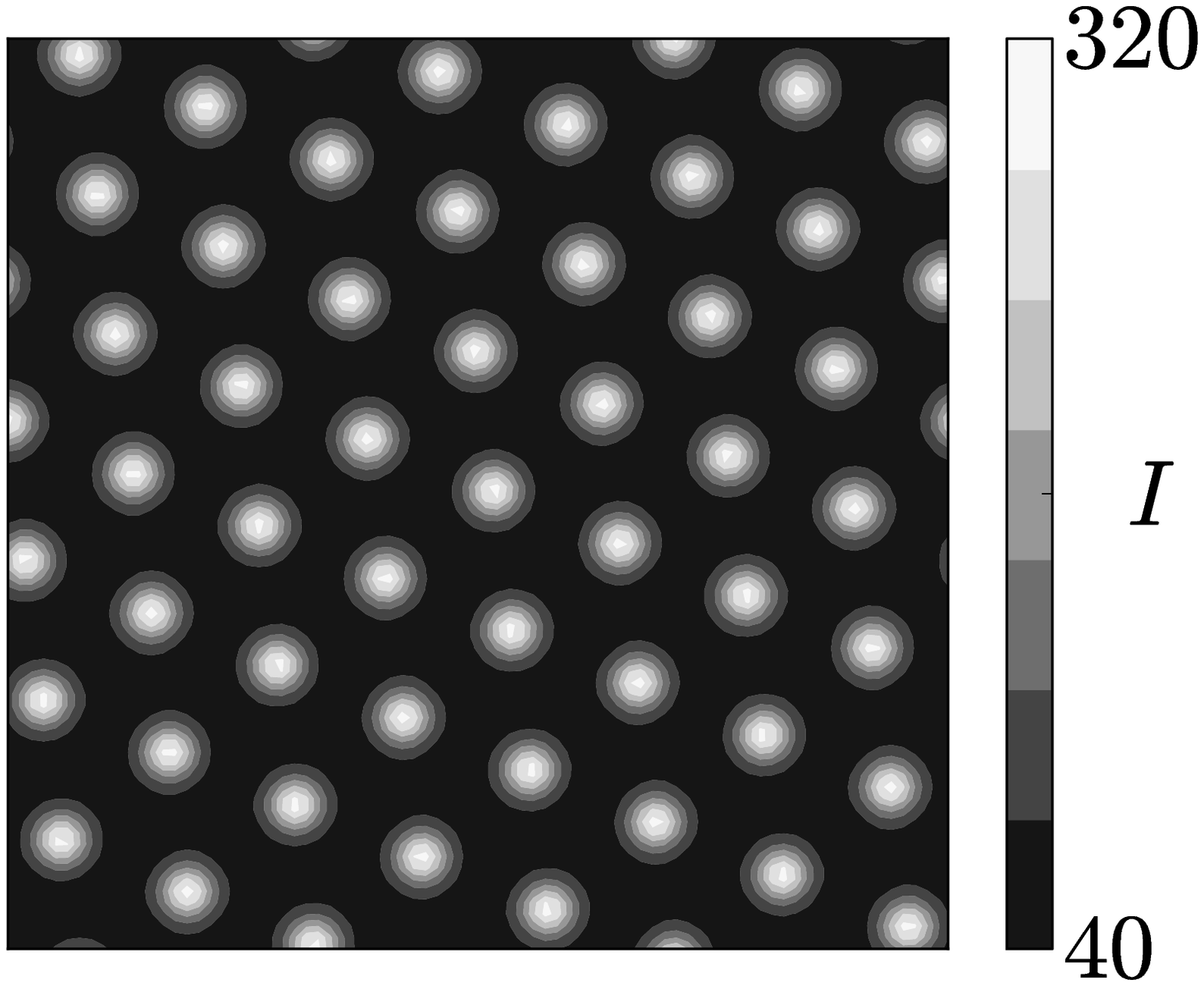}\quad
 \includegraphics[scale=0.17]{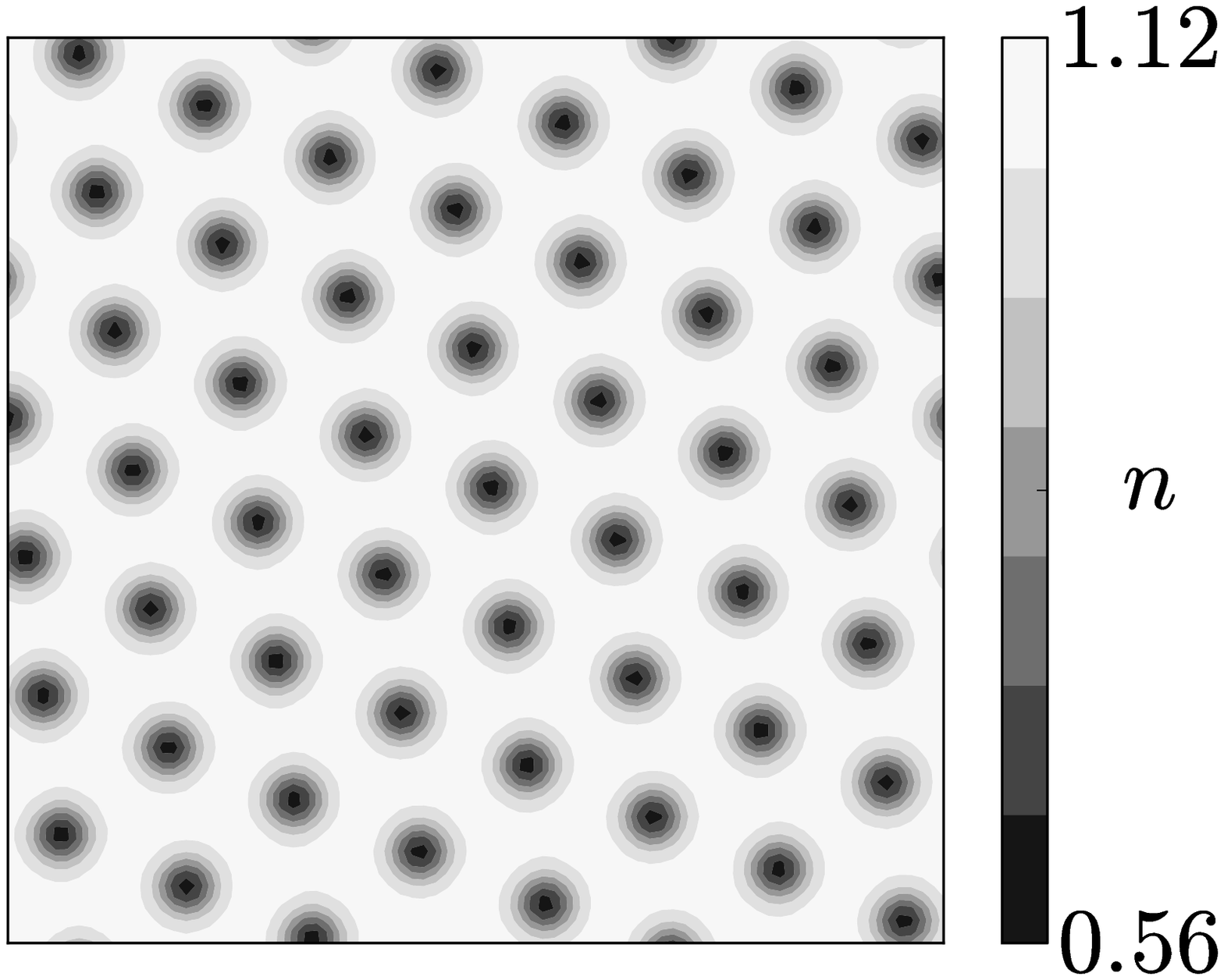}
 \includegraphics[scale=0.17]{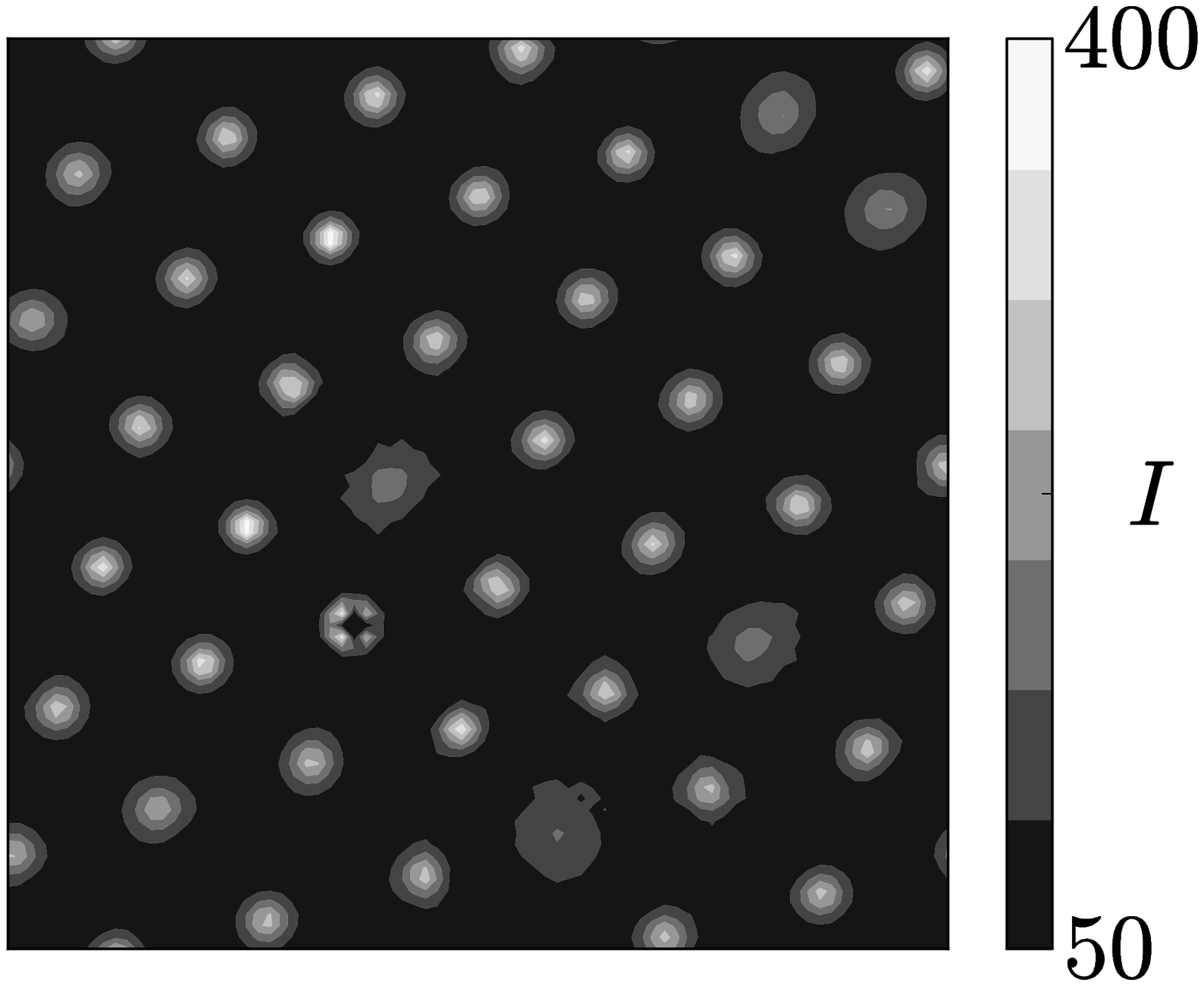}\quad
 \includegraphics[scale=0.17]{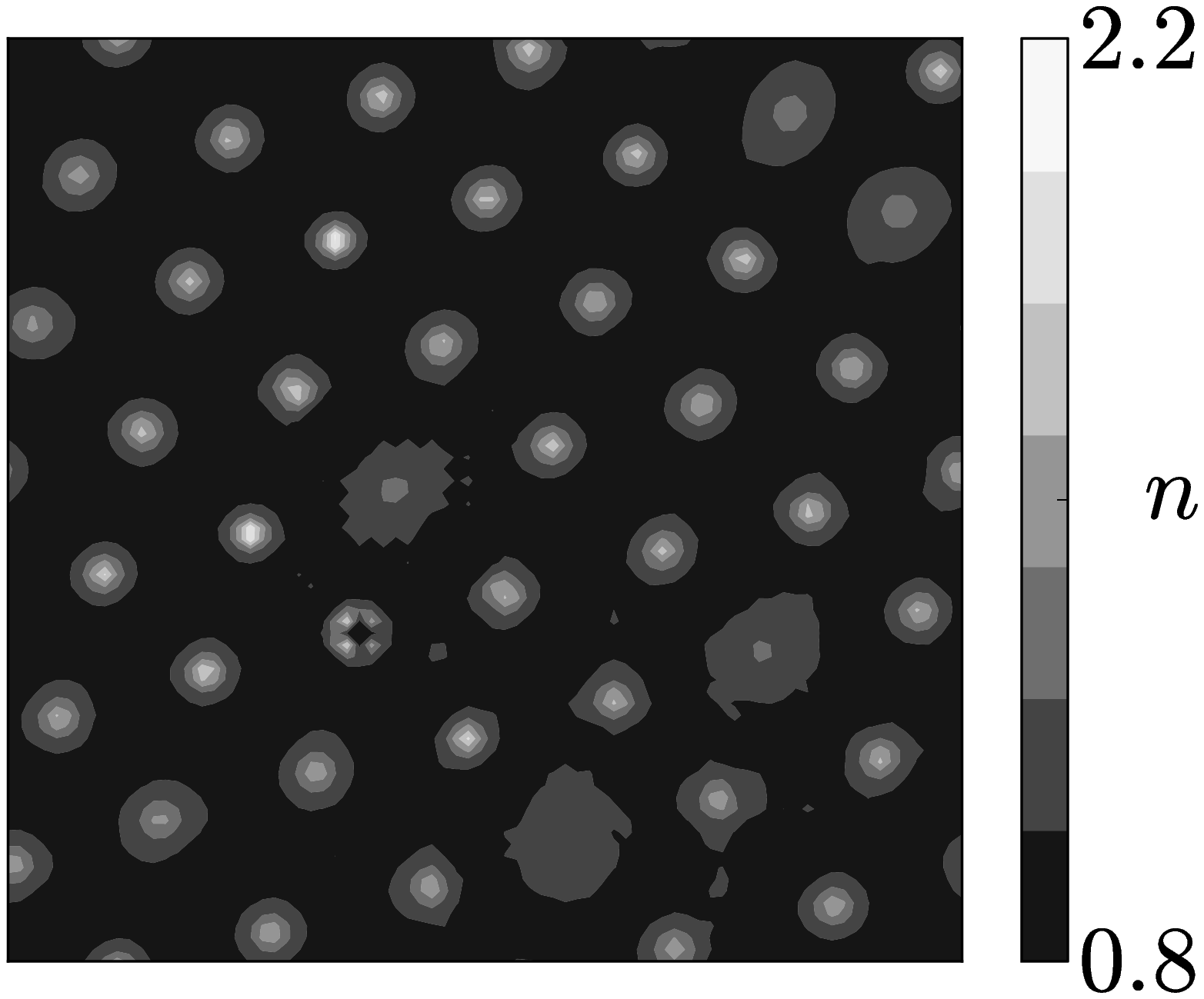}
 \caption{Opto-mechanical hexagon formation at about $7\%$ above threshold. Parameters are as in
 Fig.~\ref{fig:threshold}, with the blue-detuned case in the top panel and the red
 detuned case in the bottom panel. Field intensity is depicted on the left, the corresponding density
 modulation on the right.}
 \label{fig:simulations2D}
\end{figure}
Fig.~\ref{fig:simulations2D} presents results of numerical integration in 2D of the simple system~(\ref{eq:pureOM}), for the same parameters as Fig.~\ref{fig:threshold}, for
both signs of $\sigma$ and $I=95$, about $7\%$ above threshold.
The opto-mechanical nonlinear mechanism
leads in both cases to the formation of hexagonal structures, as is usual in systems with intensity-dependent cubic nonlinearities~\cite{Firth:PRA:92}. Note that the intensity pattern is bright hexagons
in both cases, but the density structure is honeycomb-type for blue detuning. This is because the
linear refractive index of the atom cloud is less than unity, so that a ``hole" in the cloud has relatively
high index, and so can guide light. We thus interpret the upper panels of Fig.~\ref{fig:simulations2D}
as a self-organized hexagonal network of waveguides formed by the expulsion of atoms from
the guided beams. Since the overlap of light and atoms is reduced in the fully-formed pattern the
opto-mechanical nonlinearity has a strong effective saturation, and this pattern is very stable.
For red detuning, in contrast, atoms and light attract each other, so that the light is now guided by
atomic filaments of high density (lower panels Fig.~\ref{fig:simulations2D}). As evidenced by the variation of the
amplitude of the peaks in Fig.~\ref{fig:simulations2D} (lower panel), here the interaction
is enhanced by the pattern formation, and it is perhaps unsurprising that the stability of the pattern
is much poorer for red detuning, though detailed investigation of stability issues is beyond the
scope of this Letter.\\
With a view to experimental observation of opto-mechanical pattern-forming instabilities,
we note that an effective cavity
length $L$ of a few centimeters would lead to pattern scales  of the order of
$\lambda_c\sim100\,\mu$m, requiring beam diameters of at least $\sim500\,\mu$m for the development of a well-defined transverse structure.  For standard Rb molasses at temperatures
of $T\sim100\,\mu$K the threshold intensity is of order $100\,$mW/cm$^2$, so that
2~mW of intra-cavity  power should be ample to generate opto-mechanical hexagons.
\par
% ===========================
% ====== CONCLUSIONS ========
We have investigated a mechanism for transverse optical pattern formation dominated by density
modulation effects due to dipole forces exerted by light on a cold atomic medium. By considering
a very simple ring-cavity configuration we have been able to isolate and highlight the role and
importance of opto-mechanical effects in light-atom interaction at low temperatures.
The threshold and the spatial scales of the resulting
patterns have been confirmed by numerical simulations in one and two transverse dimensions.
The required atomic temperatures and optical intensities are well within experimental
capabilities, and indeed we suggest that transverse density modulation effects may well be
present in previous studies involving more complex configurations. Our simulations predict the
formation of self-organized hexagonal wave-guiding filaments, the filaments being atom-poor or
atom-rich depending on the optical frequency.\ The resulting coupled light-matter structure can be interpreted as a self-organized and self-loaded optical lattice, optical lattices being the workhorse for applications of cold atoms solid-state physics and quantum information. As for other systems which form hexagonal patterns,
we would expect the existence of stable single filaments, i.e.\ localized states or dissipative solitons.
\begin{acknowledgments}
Financial support from the Leverhulme Trust (research grant F/00273/0) and the Engineering and Physical Sciences
Research Council (for GRMR - grant EP/H049339) is gratefully acknowledged.
\end{acknowledgments}

\end{document}